\documentclass{b98proc}
\include{psfig}

\def\Journal#1#2#3#4{{#1} {\bf #2}, #3 (#4)}
\def\NPB{{\em Nucl. Phys.} B}
\def\PLB{{\em Phys. Lett.} B}
\def\PRL{\em Phys. Rev. Lett.}

\begin{document}

\title{ DISPERSION RELATIONS IN $\bar KN$ -- $\pi Y$ COUPLED 
CHANNEL ANALYSES }

\author{Paolo M. Gensini}
\address{Dip. Fisica, Universit\`a, and Sez. I.N.F.N., Perugia, Italy}
\author{Rafael Hurtado}
\address{Dip. Fisica, Universit\`a, and Sez. I.N.F.N., Perugia, Italy, and 
C.I.F., Bogot\'a, Colombia}
\author{Galileo Violini}
\address{Dip. Fisica, Universit\`a della Calabria, and Gruppo Coll. I.N.F.N., 
Cosenza, Italy, and U.N.E.S.C.O., Santiago, Chile}

\maketitle

\abstracts{Rapidly convergent, unsubtracted dispersion relations have been 
used to test coupled--channel analyses of low--energy $S=-1$ meson--baryon 
channels.}

\par Processes like $\pi\Lambda$ and $\pi\Sigma$ elastic scattering, or 
$\pi\Lambda\to\pi\Sigma$, are not accessible to experiment: their 
amplitudes could only be accessed as a byproduct of the coupled--channel 
{\bf K}--matrix formalism employed to analyse $\bar KN$ interactions. 
Our purpose is to analyse in detail the dispersion relations that can 
be written for these channels and saturated with a minimum of additional 
hypotheses, in order to understand whether analyticity constraints, imposed 
on these channels, combined with the results of suitable experiments at 
DA$\Phi$NE, might improve our knowledge on the $\bar K N$ -- $\pi Y$ 
coupled--channel amplitudes. The experiments used to determine the {\bf 
K}--matrix elements had statistics much lower than those attainable at 
DA$\Phi$NE, and this machine will be able to explore in detail the 
very-low energy region around the $\bar K^0n$ charge-exchange threshold, 
where energy--dependent effects due to the $K^-p$ -- $\bar K^0n$ mass 
difference are expected to show up.

We use the low--energy amplitudes obtained in a coupled--channel, {\bf 
K}--matrix analysis \cite{Kim} to calculate the $\pi Y\Sigma$ couplings 
by conventional dispersion relations (D.R.'s) applied to the three $\pi Y 
\to \pi Y'$ reactions, namely elastic $\pi\Lambda$ and $\pi\Sigma$ 
scattering and $\pi\Lambda\to\pi\Sigma$. 
We have chosen amplitudes with the fastest decrease as $\omega\to\infty$: 
in terms of the invariant amplitudes $A(\omega,t)$, $B(\omega,t)$ and 
$C(\omega,t)$ (with $C(\omega,t) = A(\omega,t) + \omega 
B(\omega,t)/[1-t/(M+M')^2]$ being the amplitude whose imaginary part in 
the forward direction is given by the optical theorem as 
$k\sigma_{tot}(k)/{4\pi}$), we use $B(\omega,0)/\omega$ for the elastic 
$\pi\Lambda$ and $\pi\Sigma$ scattering, and $C(\omega,0)/\omega$ for 
the inelastic process $\pi\Lambda\to\pi\Sigma$. 
$A$ and $C$ are dominated by the S--wave, while $B$ by the P--waves: 
since the most recent K--matrix analyses are purely S--wave, 
and the existence of the $\Sigma$(1385) requires at least inclusion of 
the P$_{13}$ one as well, a calculation using such parametrizations 
would be useless for $B$, whereas for $A$ and $C$ would lead to 
determine an ``effective coupling'' for both the $\Sigma$ and the 
$\Sigma$(1385) \cite{Mar}. 
Due to the absence of information on the I = 2 $\pi\Sigma$ amplitude, in 
the elastic $\pi\Sigma$ DR we are forced to use \cite{Cha} the crossing-even 
combination of isospin amplitudes $2B_1(s,t,u)-B_0(s,t,u)$, 
which is the only one free of the I = 2 part both in the $s$ and $u$ 
channel.

\begin{figure}[ht]
\psfig{figure=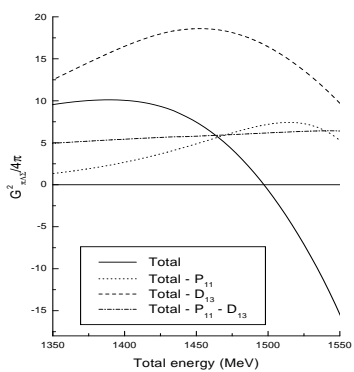,height=6cm,width=5cm}
\caption{$G_{\pi\Lambda\Sigma}^2/4\pi$ from 
$\pi\Lambda\to\pi\Lambda$}
\label{f1}
\end{figure}

For the low--energy region there exist two {\sl published} {\bf K}--matrix 
parametrizations extending down to the $\bar KN$ threshold and including 
S--, P-- and D--waves: the second (Gupta, {\sl et al.} \cite{Gup}) includes 
also F--waves and is an amalgamation of different analyses, but it does not 
reproduce the expected structures below the $\bar KN$ threshold: 
this leaves only the first one (the old Kim's parametrization of 1967 
\cite{Kim}) as the only viable choice for such calculations. 
We have slightly corrected Kim's D--waves in order to describe with correct 
threshold behaviours the very--low energy region: for the 
D$_{03}$ wave we have limited ourselves to modify the interaction--radius 
dependence so as to eliminate a spurious singularity in the unphysical 
region, replacing $(X^2+k^2)^{-2}$ with $(X^4+2X^2k_0^2+k^4)^{-1}$ ($k_0$ 
is the $\bar K N$ c.m. momentum at the energy of the $\Lambda(1520)$ 
resonance), while for D$_{13}$ we have replaced the constant scattering 
length approximation used in Kim's fit by a rank--zero {\bf K}--matrix 
reproducing Kim's partial amplitudes around the D--wave resonance.

\begin{figure}[ht]
\psfig{figure=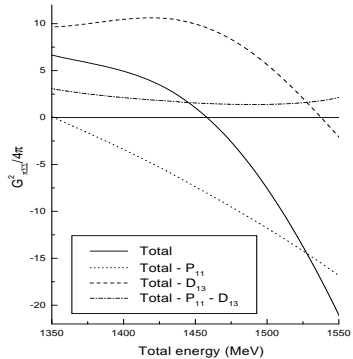,height=6cm,width=5cm}
\caption{$G_{\pi\Sigma\Sigma}^2/4\pi$ from 
$\pi\Sigma\to\pi\Sigma$ and $\pi\Lambda\to\pi\Lambda$}
\label{f2}
\end{figure}

\begin{figure}[ht]
\psfig{figure=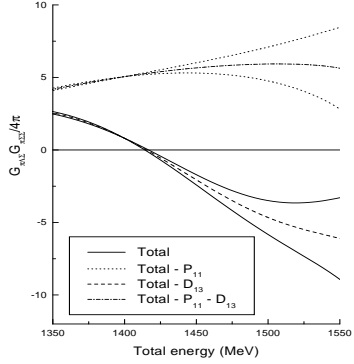,height=6cm,width=5cm}
\caption{$G_{\pi\Lambda\Sigma}G_{\pi\Sigma\Sigma}/4\pi$ from 
$\pi\Lambda\to\pi\Sigma$}
\label{f3}
\end{figure}

We present here the results of this analysis (where for completeness we 
include also estimates of the contributions from intermediate-- and 
high--energy regions to the dispersive integrals, which turn out however 
to be quitesmall) as the values for the different couplings versus the 
energy at which the D.R.'s are evaluated, both with and without the (small) 
P$_{11}$ and D$_{13}$ waves. For elastic $\pi Y$ scattering the results 
without the last one correspond to the old evaluations performed by Chan and 
Meiere \cite{Cha}, but only at the energy of the $\bar KN$ threshold. 
In the elastic $\pi Y$ scattering (figg.~\ref{f1},~\ref{f2}) 
results there is an indication of large 
cancellations between the two waves (and therefore an essential instability 
for an otherwise unconstrained analysis as Kim's one). Being the coupling 
of the D$_{13}$ wave to the $\pi\Sigma$ channel quite small, this 
cancellation does not materialise in the inelastic $\pi\Lambda\to\pi\Sigma$ 
channel D.R. (fig.~\ref{f3}), which therefore is 
extremely sensitive to the P$_{11}$ contribution. 
This instability does not allow to determine the $G_{\pi Y\Sigma}$ coupling 
constants with the accuracy claimed by previous calculations \cite{Cha}. If 
the (combined) contributions from P$_{11}$ and D$_{13}$ waves were much 
overestimated, then it would still be possible for these couplings to 
agree with each other and with SU(3)--symmetry predictions.

These results show that the low--energy $S=-1$ meson--baryon 
sector is less well known than currently believed, and that it deserves 
more accurate experiments and analyses if one wants to reach 
a level of knowledge comparable to that of the $\pi N$ one 
\cite{Gen1,Gen3,Hur}. DA$\Phi$NE is the right machine to achieve this goal.

Also, much more stringent constraints from unitarity, analyticity and 
crossing symmetry than those from $\bar KN$ forward D.R.'s (as {\sl e.g.} 
used by A.D.~Martin) have to be enforced if a stable and accurate 
parametrization of the low--energy region has to be extracted from the 
data \cite{Gen2}.

\end{document}